# Political elections and uncertainty - Are BRICS markets equally exposed to Trump's agenda?

Jamal Bouoiyour [†] and Refk Selmi [‡] [§]

**Abstract:** There certainly is little or no doubt that politicians, sometimes consciously and sometimes not, exert a significant impact on stock markets. The evolving volatility over the Republican Donald Trump's surprise victory in the US presidential election is a perfect example when politicians, through announced policies, send signals to financial markets. The present paper seeks to address whether BRICS (Brazil, Russia, India, China and South Africa) stock markets equally vulnerable to Trump's plans. For this purpose, two methods were adopted. The first presents an event-study methodology based on regression estimation of abnormal returns. The second is based on vote intentions by integrating data from social media (Twitter), search queries (Google Trends) and public opinion polls. Our results robustly reveal that although some markets emerged losers, others took the opposite route. China took the biggest hit with Brazil, while the damage was much more limited for India and South Africa. These adverse responses can be explained by the Trump's neo-mercantilist attitude revolving around tearing up trade deals, instituting tariffs, and labeling China a "currency manipulator". However, Russia looks to be benefiting due to Trump's sympathetic attitude towards Vladimir Putin and expectations about the scaling down of sanctions imposed on Russia over its role in the conflict in Ukraine.

**Keywords:** US presidential election; Trump's agenda; stock markets; BRICS; event study; social media; search queries; public opinion polls.

[†] CATT, University of Pau, France.
[‡] University of Tunis, Tunisia; University of Pau, France.
[§] Corresponding author : *jamal.bouoiyour@univ-pau.fr*
Full Address: Avenue du Doyen Poplawski, 64016 Pau Cedex, France; Phone: 33 (0)5 59 40 80 01; Fax: 33 (0)5 59 40 80 10

## 1. Introduction

Major stock indices around the world witnessed huge negative fluctuations during the initials hours after Donald Trump won the US presidential election. The MSCI Emerging Markets index collapsed markedly since 9 November 2016. Among emerging markets, the BRICS (Brazil, Russia, India, China and South Africa) shares fell significantly. In a weekly report published by the Brazil's Central Bank on Monday 14 November 2016, the GDP retraction for 2016 has been revised from 3.31 percent to 3.37 percent. In fact, the São Paulo's stock market index fluctuated extremely since the announcement of the US election outcome. After decreasing by about 0.98 percent early in Wednesday, 09 November morning, the index bounced back by 0.11 percent in the afternoon. Since the event day, the MSCI India stock index fell by 7.1 percent compared to 4.9 percent drop in emerging markets more broadly; also, the rupee lost more than 3 percent against the dollar. Likewise, China's stock index dropped significantly (i.e., the Shangai index plunged as much as 3.6 percent on 9 November by mid-afternoon) on fears that President-elect Trump's protectionist proclivities will harm their trade and then exacerbate the current Chinese economic slowdown. For Trump, China's manufacturing hub and low-cost production have threatened the US economy. Besides, South Africa's share index tumbled as much as 4 percent on Wednesday, but it rebounded very slightly in the end of the day given the surge in gold miners as traders and investors search generally for safe havens under uncertain period. In general, emerging markets struggled as the rally in the dollar following the Trump's triumph dampened demand for emerging market assets. If there is one country viewed gaining from political risk from Washington, it's Russia. Unlike the rest of BRICS equities, the Market Vectors Russia (RSX) exchange traded fund rose by 4.5 percent, beating the S&P 500 and the MSCI Emerging Markets Index.

Even though many analysts and one recent research (Bouoiyour and Selmi 2016 b) are talking about how US stock markets might react to Trump's win, there are a number of other countries that saw their markets respond significantly after the polls close. The Trump's America First protectionist plans may hurt heavily the emerging markets including the BRICS. Indeed, Trump claimed the abandon of the tariff-cutting Trans Pacific Partnership trade agreement between the U.S. and particular emerging nations. Although it is still unclear whether or not Trump's promises will translate into actual economic and political policies, market participants appear to be concerned by ongoing volatility because of the Trump's protectionist rhetoric. All BRICS leaders aim, undoubtedly, to promote economic growth and curtail foreign capital flight while controlling for political turmoil and overcoming the harmful protectionism consequences. The International Monetary Fund (IMF2016) anticipated that a rise in global protectionism could decrease the global GDP by more than 1.5 percent over the next years. But it might be relevant for market participants to differentiate between the countries best able to weather the storm, and those unable to avoid the adverse effects of uncertainty surrounding the Trump's economic agenda.

Given these considerations, the present study point out the prominence of answering some critical questions. What Trump's election victory means for BRICS shares? Do BRICS move from markets' strength to vulnerability? Are BRICS stock markets equally exposed to Trump's plans? To address these questions, we use the standard market model event study methodology as originally described by Dodd and Warner (1983) and Brown and Warner (1985). The event study aims at investigating the average stock market response to a specific stock market event (here the announcement of Trump's victory on 08 November 2016). Beyond the analysis of the effect of the day relative to the announcement of Trump winning on the abnormal returns, this study also assesses how respond the BRICS stock returns to vote intentions based social media (Twitter), search queries (Google Trends) and polling data as indicators of public interest-levels in the US presidential election.

In doing so, we unambiguously document that the BRICS stock markets are heterogeneously exposed to Trump's stunning triumph. While some markets emerged losers (China, Brazil, India and South Africa, in this order), others appeared winners (Russia). The victory of Donald Trump is viewed as detrimental for BRICS markets especially because of Trump's protectionist rhetoric. However, a potential factor which makes investors more bullish toward Russian shares is the possible easing of the western sanctions regime against Russian companies.

The outline of the paper is as follows: Section 2 includes a brief discussion of the theory on the effect of political uncertainty on financial markets. Section 3 outlines the methodology adopted and describes the data. Section 4 reports and discusses the main findings. Section 5 concludes.

## 2. Political elections, uncertainty and financial markets: Some theoretical considerations

The interest in examining the relationship between stock markets and political uncertainty is among researchers for a long time. In general, the political risk is associated with heavier stock return volatility. Normally, the stock markets have to make important choices based on the expected future economic policy decisions of the new government and the resulting policy circumstances (Brogaard and Detzel 2015; Schiereck et al. 2016; Bouoiyour and Selmi 2016 a). Such policy changes put downward pressure on stock prices, particularly if the uncertainty is extreme (Pastor and Veronesi 2012). Once the political uncertainty become less pronounced, stock prices would bound back (Pantzalis et al. 2000). But some events may have persistent effects. For instance, in the case of Brexit, the uncertainty is likely to still higher until it becomes clearer what the future relations among the UK and the European Union will be, continuing to exert a harmful influence on stock prices (Bouoiyour and Selmi 2016 b; Schiereck et al. 2016).

Even though political uncertainty takes various shapes and forms including changes in the government and changes in the domestic and foreign policies, the present research focuses on one kind of political uncertainty, which is associated with elections. The latter constitute a major event for re-distribution of political power, which may have meaningful implications for the future political and economic prospects of a country. There is considerable debate regarding the impact of elections on asset price variation (Kim and Mei 2001; Akmedov and Ekaterina 2004; Canes-Wrone and Jee-Kwang 2014; Bouoiyour and Selmi 2016 a). Nevertheless, there is a large consensus that political uncertainty makes financial markets extremely volatile, particularly after close elections or in response to election results that may lead to radical policy changes (Canes-Wrone and Jee-Kwang 2012). There are at least three reasons that election may exacerbate the financial market volatility. Firstly, a potent political uncertainty surrounding the election outcome may intensify the asymmetries between informed and uninformed market participants. Secondly, the deeper uncertainty over the US presidential election may amplify the ambiguity across market participants about economic fundamentals influencing the share values. The Trump's storming victory has ramped up uncertainty over the policies he will pursue. Several analysts proclaimed that the only certainty about US President-elect Donald Trump's incoming administration is the uncertainty that will attend it. This is seemingly true with regard to a main sensitive policy area. If Trump administration cut taxes and undertake a massive infrastructure program, America's budget deficits will increase substantially. This accompanied with the Federal Reserve's gradual interest-rate hikes will appreciate the dollar, and deteriorate the so-called emerging-market currencies, and shift money from the rest of the world to the US. This is viewed as a very anxious -if not terrifying- prospect. Thirdly, the political uncertainty in election may disrupt the normal functioning of financial markets since the Trump's proclamations on different topics (the withdrawal from NAFTA, the renegotiation free-trade agreements resulting more isolated and less open US markets) remain conditional to the overall congress opinion and the legal challenges from private firms which may play a pivotal role in deterring Trump's administration from implementing these measures (Bouoiyour and Selmi 2016 a). All of these considerations may be of utmost relevance for “politically sensitive” industries, i.e., the companies whose economic fortunes are more likely to be significantly influenced by political continuity or discontinuity.

The literature has put much effort in refining measures of uncertainty (Bloom 2009; Bloom et al. 2012; Cesa-Bianchi et al. 2014; Jurado et al. 2015, etc…). In general, uncertainty is defined as the conditional volatility of a disturbance that is unforecastable. A challenge in empirically analyzing the uncertainty and its dependence to other macroeconomic and financial phenomena is that no objective measure of uncertainty exists. Throughout the rest of our study, we analyze the uncertainty over US political elections via two dimensions: (a) the way in which the 2016 US presidential election was communicated by media and social networking and the public opinion polls; and (b) the time leading up to an election or the time of government transition after the election by using a dummy variable for the day relative to the announcement of election result.

## 3. Methodology and data

To quantify the effects of Trump’s victory on BRICS stock markets, we conduct two methodological steps. On the one hand, we analyze the impact of the 2016 US election event (a dummy variable that takes a value 1 on 08 November 2016 and 0 otherwise) on the BRICS abnormal returns. On the other hand, we assess the impact of the vote intentions on the BRICS stock returns. Precisely, we offer a new approach to identify the peoples’ opinions about Trump’s win by using data from social media, search queries and public opinion polls.

### 3.1. The event study methodology

The event study methodology, first proposed byWarner (1983) and Brown and Warner (1985), is designed to examine the impact of a specific event on a dependant variable. A commonly used dependent variable in event studies is the stock price. Accurately, an event study is an analysis of the changes in stock price beyond expectations (Abnormal returns) during a precise period of time (event window), such as the abnormal returns are attributed to the onset of such event. Overall, the purpose is to test if there is an abnormal stock price effect associated with an event. We define the day "0" as the day of the event for a given stocks. Thereafter, the estimation and event windows can be determined (see Figure 1). The interval [T1+1, T2] is the event window with length L2=T2-T1-1, whereas the interval [T0+1, T1] is the estimation window with length L1=T1-T0-1. The length of the event window often depends on the ability to accurately date the announcement date. If one is able to date it precisely, the event window will be less lengthy and capturing the abnormal returns will be more adequate.

**Figure 1. Event study windows**

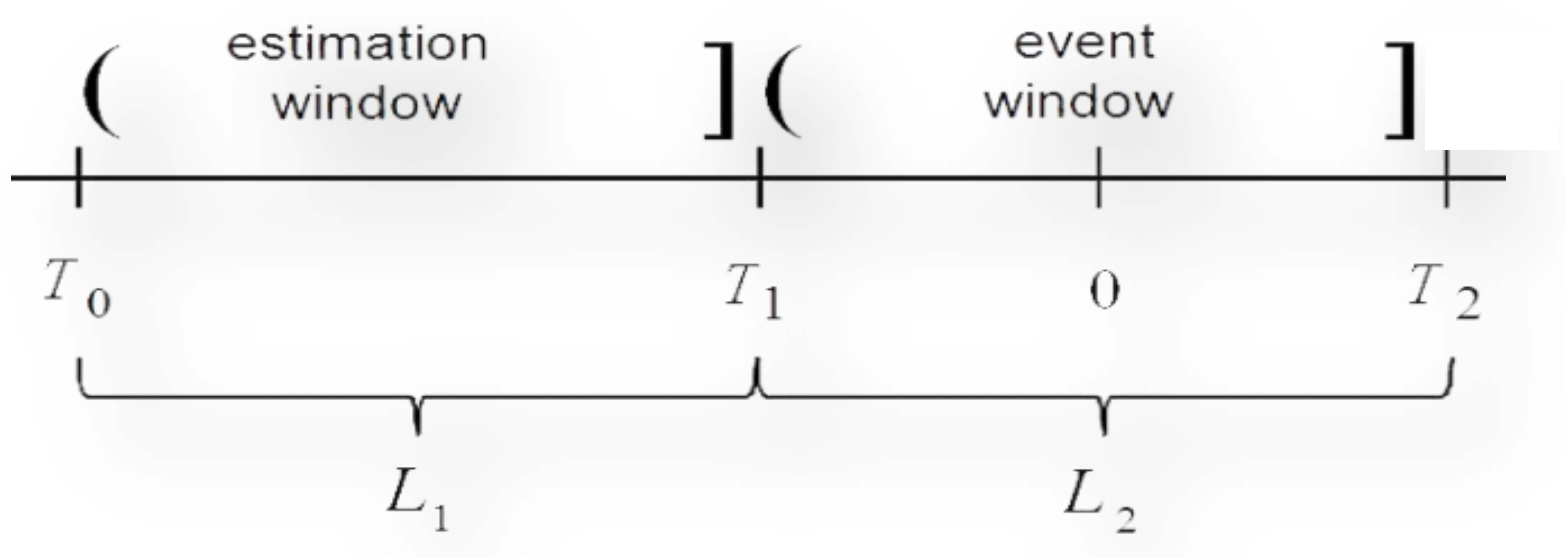

For our case of study, we use for each BRICS equity a maximum of 120 daily stock returns observations for the period around the ultimate election result, beginning at day - 110 and ending at day + 5 relative to the event. The first 110 days (- 115 through -5) is denoted as "the estimation period", and the following 11 days (- 5 through + 5) is designated as "the event period". The cumulative abnormal return (CAR) for a sector *i* during the event window [ $\tau_1$ ; $\tau_2$ ] surrounding the event day $t$ = 0, where [ $\tau_1$ ;$\tau_2$ ] = ∈ [ −5 ;+5] , is expressed as follows:

$$CAR_{i,[\tau_1,\tau_2]} = \sum_{t=\tau_1}^{\tau_2} (R_{i,t} - \hat{\alpha}_i - \hat{\beta}_i R_{M,t}) \qquad (1)$$

where $CAR_{i,[\tau_1,\tau_2]}$ is the cumulative abnormal return of share *i* during the event window [$\tau_1; \tau_2$], $R_{i,t}$ is the realized return of stock *i* on day *t*[4], $R_{M,t}$ is the return of the benchmark index of stocks *i*, $\hat{\alpha}_i$ and $\hat{\beta}_i$ are the regression estimates from an ordinary least squares (OLS) regression for 110 trading day estimation period until t = −5. We utilize the *MSCI* emerging stock market return as the benchmark index. We set our event day for the Trump's win event to 8 November 2016.

Then, an OLS regression of the observed cumulative abnormal return for each BRICS shares on the announcement day of the Trump winning is estimated. For this purpose, we use daily data for the stock market indices of Brazil's Ibovespa, China's Shanghai index, Russia's RTS index, the India's NSE and South Africa's FTSE/JSE.

The equation to estimate is denoted as:

$$CAR_{i,[\tau_1,\tau_2]} = \delta_0 + \delta_1 Event + \varepsilon_i \quad (2)$$

where $CAR_{i,[\tau_1,\tau_2]}$ is the cumulative abnormal returns (the dependent variable), *Event* is a dummy variable which takes the value of 1 on the day of the US election outcome and 0 otherwise, and $\varepsilon_i$ is the error term.

Another objective of this research is to see whether the event-study findings are sensitive to the inclusion of potential control variables. Generally, major global financial and economic factors could be channels through which fluctuations in the world's economic and financial conditions are transmitted to BRICS stock markets. These factors include the West Texas Intermediate (*WTI*) oil price, the world gold price (*Gold*) and the silver price (*Silver*). The *WTI* has been widely used in the literature as the benchmark price for global oil markets. The *WTI* crude oil is among the most traded oil on the world markets, and therefore is significantly affected by macro-financial variables. Due to their surges under uncertain circumstances, the precious metals (gold and silver) have been perceived as a hedge against sudden shocks and also a safe haven over extreme stock market fluctuations

[4] Daily stock returns are calculated as the first natural logarithmic difference of the stock price.

(Baur and Lucey 2010; Hood and Mallik 2013). According to Baur and McDermott (2010), we characterize safe havens by their negative and significant correlations with asset markets during financial turmoil or troubled times. In addition, the Bitcoin[5]'s considerable climb alongside the announcement of Trump's victory has led to affirm its validity as a safe haven investment. As a reaction to the uncertainty surrounding the US election result, the asset markets around the world plunged as investors were concerned about ongoing volatility. This has yielded to a trend towards questioning the effectiveness of standard economic and financial structures which govern the conventional monetary and financial system. Here, the digital currency is leading the charge by providing a completely decentralized secure alternative to fiat currencies during times of economic and geopolitical unrest. The *WTI*, *Gold* and *Silver* prices data are sourced from DataStream of Thomson Reuters, while the Bitcoin price data in US dollars are collected from CoinDesk at www.coindesk.com/price. The variables under study were transformed by taking natural logarithms to correct for heteroskedasticity and dimensional differences.

The function to estimate is expressed as follows:

$$CAR_{i,[\tau_1,\tau_2]} = \chi_0 + \chi_1 Event + \chi_2 WTI + \chi_3 Gold_t + \chi_4 Silver_t + \chi_5 Bitcoin_t + \vartheta_i \quad (3)$$

where $CAR_{i,[\tau_1,\tau_2]}$ is the cumulative abnormal returns and $\upsilon_i$ is the error term.

### 3.2. A regression-based intention votes

Unlike the event-study methodology based on regression estimation of abnormal returns that helps to answer whether BRICS equities uniformly respond to the announcement of Trump winning, in this section, we introduce the concept of internet concern as quantitative measure to test if extracting public moods related to

[5]Bitcoin was created in 2009 by an anonymous programmer under the pseudonym Satoshi Nakamoto and has since achieved a widest level of international recognition. Unlike the fiat currencies, Bitcoins are digital coins which are decentralized, not issued by any government or legal entity and not redeemable for gold or any other commodity. Bitcoins rely on cryptographic protocols and a distributed network of users to mint, store, and transfer. Instead, investors perform their business transactions themselves without any intermediary. The peer-to-peer network eliminates the trade barriers and makes business easier (Bouoiyour et al. 2016).

US election exerts a significant influence on BRICS stock markets. Millions of people daily interact with search engines, creating valuable sources of data regarding the 2016 US election. In brief, the Internet search seems a potential tie allowing analyzing the public opinions towards the election.

Recent literature evaluated how online information predicts Grexit (Mitchell et al. 2012; Bouoiyour and Selmi 2015, among others) and the economic and financial costs of Brexit (Bouoiyour and Selmi 2016 b). We attempt, in the following, to demonstrate that social media discussion and search related queries for the 2016 US election help us to track the evolution of markets' beliefs about US presidential election outcome. Twitter is becoming very popular among financial professionals. It permits them to comment on economic and political events and to distribute their view to either their followers or even a wider audience in an extremely speedy way. Many people use their Twitter accounts to express and disseminate their opinions on the US election. The advantage of using Twitter data for research purposes is that (1) users not only receive information but can actively share information, (2) tweets can be used to extract not only a consensus view on such event, but also the degree of agreement or disagreement.

A further task of this study is to use public opinion polls to measure the intention votes toward Trump. The pollsters' reports and press releases often start with asking a specific question and then present tables with the statistical proportions of poll respondents giving all the answers. For the case of US presidential election, the question was: "If the general election were held today, and the candidates were Hillary Clinton for the Democrats and Donald Trump for the Republicans, for whom would you vote? If not sure, or would not vote, ask: Toward which do you lean?" The polls report the results used here to explain the variation of BRICS stock returns.

In brief, OLS regressions of the stock market return (*STR*) for each BRICS country on three intention votes' indicators (Google Trends, Twitter searches and polling data transformed in log) are estimated. *STR* is calculated by considering the ratio stock price (in log) at time t and the lagged stock price, i.e., $STR_t = \log(\frac{P_t}{P_{t-1}})$ where $P_t$ is the stock price.

$$STR_t = \lambda_0 + \lambda_1 GoogleTrends_t + \upsilon_i \quad (4)$$

$$STR_t = \theta_0 + \theta_1 Twitter_t + \iota_i \quad (5)$$

$$STR_t = \delta_0 + \delta_1 polls_t + \tau_i \quad (6)$$

where $\upsilon_i, \iota_i$ and $\tau_i$ are the error terms.

To avoid possible methodological bias regarding omitted variable, a vector of additional explanatory variables (discussed above) is incorporated in the models (4), (5) and (6). Precisely, we estimate the following equations:

$$STR_t = \eta_0 + \eta_1 GoogleTrends_t + \eta_2 WTI_t + \eta_3 Gold_t + \eta_4 Silver_t + \eta_5 Bitcoin_t + \xi_i \quad (6)$$

$$STR_t = \upsilon_0 + \upsilon_1 Twitter_t + \upsilon_2 WTI_t + \upsilon_3 Gold_t + \upsilon_4 Silver_t + \upsilon_5 Bitcoin_t + \zeta_i \quad (7)$$

$$STR_t = \beta_0 + \beta_1 polls_t + \beta_2 WTI_t + \beta_3 Gold_t + \beta_4 Silver_t + \beta_5 Bitcoin_t + \gamma_i \quad (8)$$

where $\xi_i$, $\zeta_i$ and $\gamma_i$ are the error terms.

We use daily time-series data related to the Trump and US presidential election over the period from 01/08/2015 to 31/12/2016. The search queries index for keyword “Trump win” has been retrieved from Google Trends at http://www.google.com/trends/. Note that in twitter #US election was associated with the Trump’s victory and it was not possible to retrieve keywords in twitter. Hash tags (#) were available only in twitter. The polling data were collected from Real Clear Politics at http://www.realclearpolitics.com/epolls/latest_polls/.

## 4. Discussion of results

### 4.1. Event study results

Figure 1 graphically depicts the CAR performance of BRICS stocks over the announcement of Donald Trump's win in US presidential election on 08 November 2016. It is clearer from the graphs that the BRICS stock market were not equally exposed to the US election outcome either for the day relative to the announcement of Trump's victory (t=0) or for the [−5; + 5] event window. Although all the emerging markets face evolving volatility, the Trump's unexpected triumph is likely to exert heterogeneous effects on BRICS equities. From a first look to the following chart, we can distinguish two groups of countries. The first group includes Brazil, India, China and South Africa where a sharp drop of stock values is found during the election day and over [+1; +5] event window. The second group is formed by Russia where we note a marked increase in the abnormal stock returns over [0; +5] event window).

**Figure 1. Cumulative abnormal return of BRICS stocks: [−5; +5] event window**

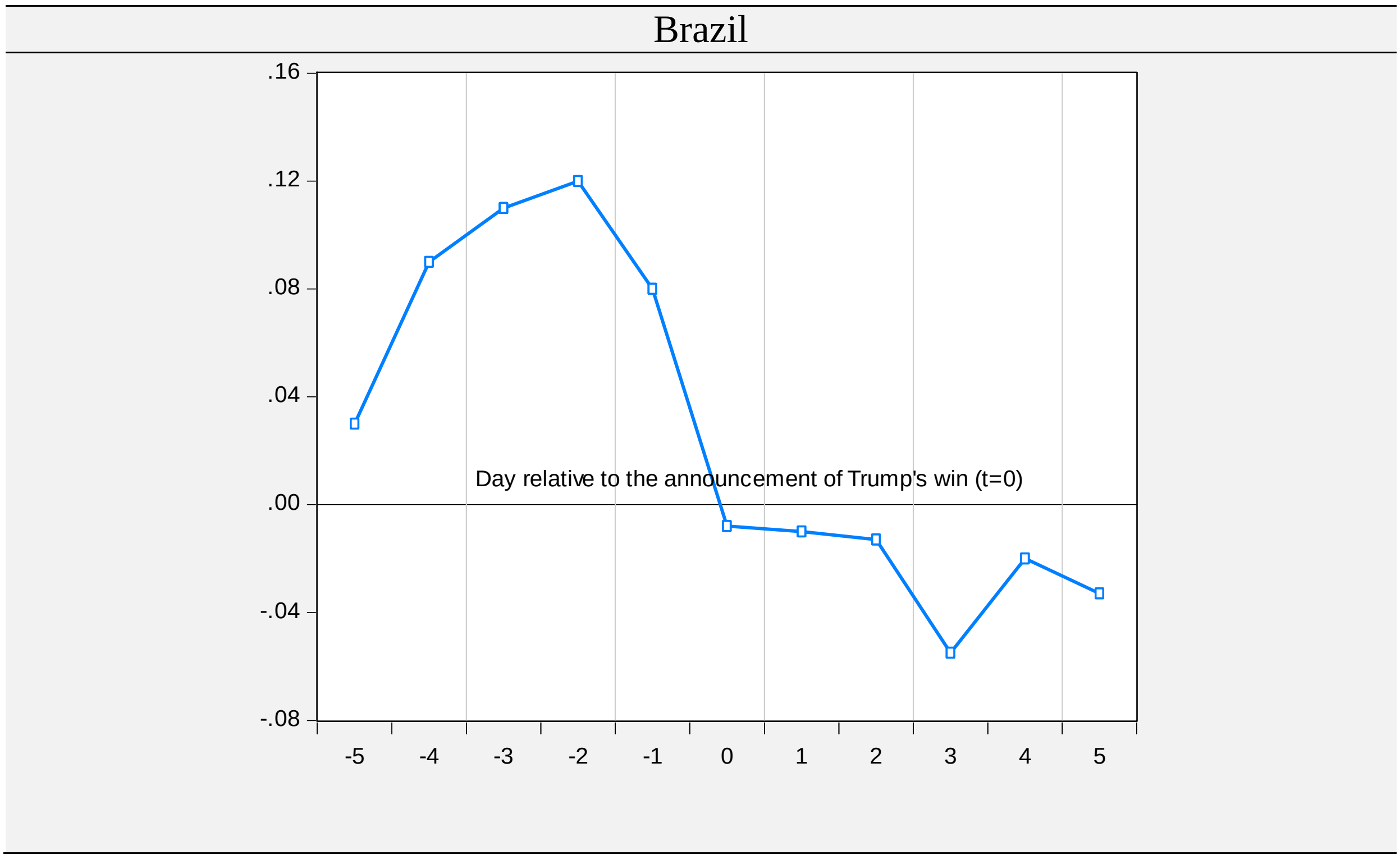

Russia

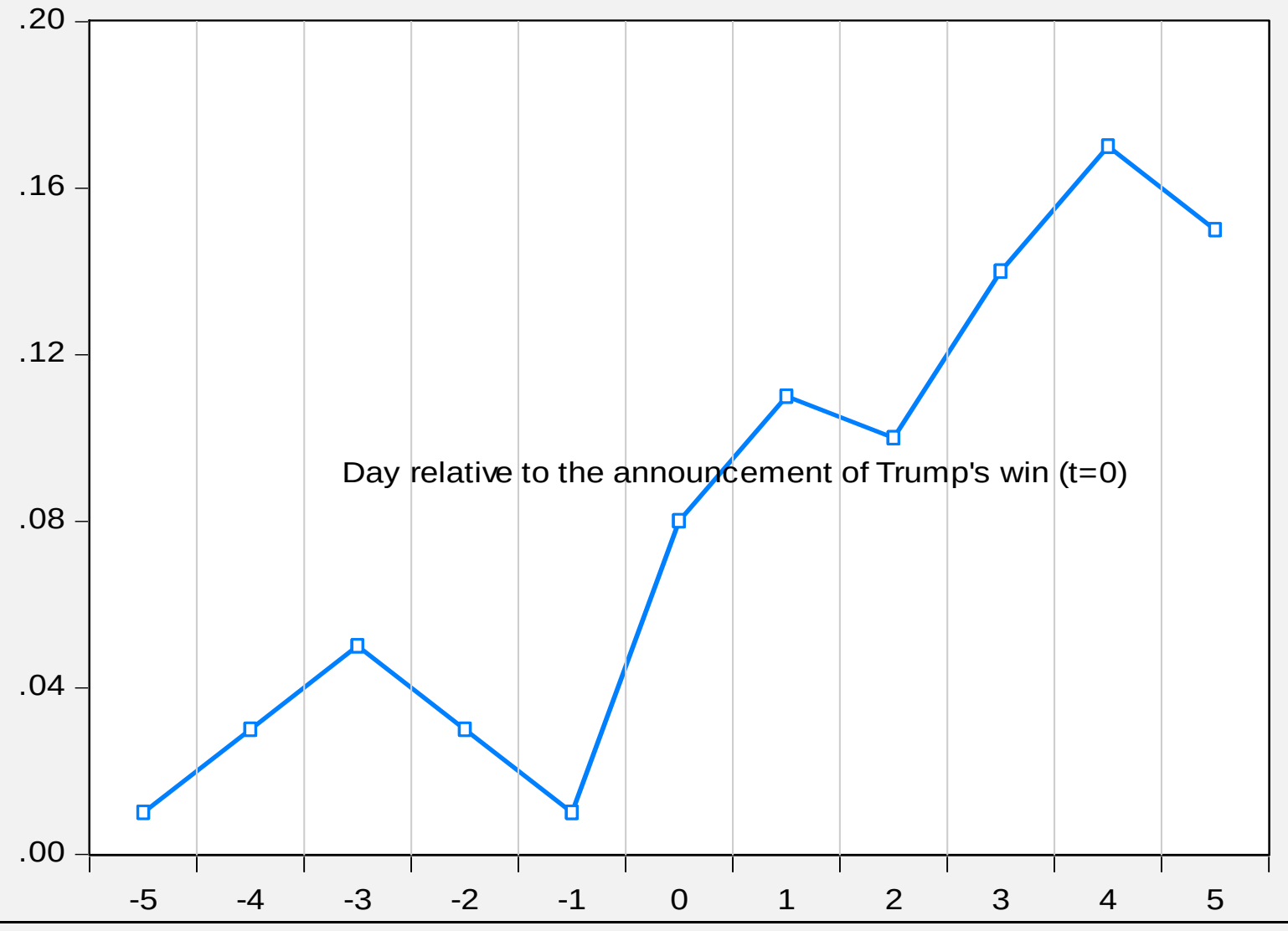

India

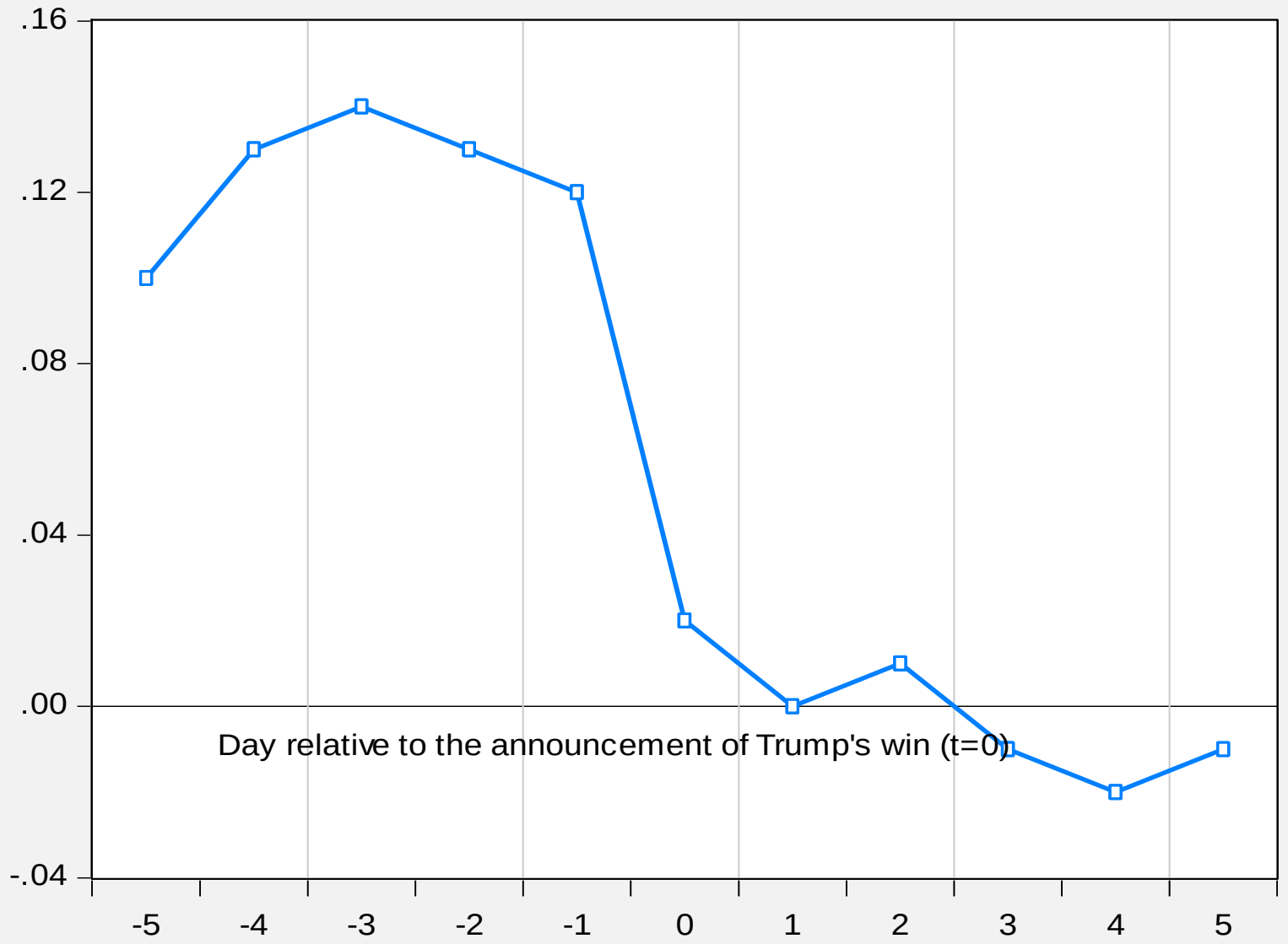

China

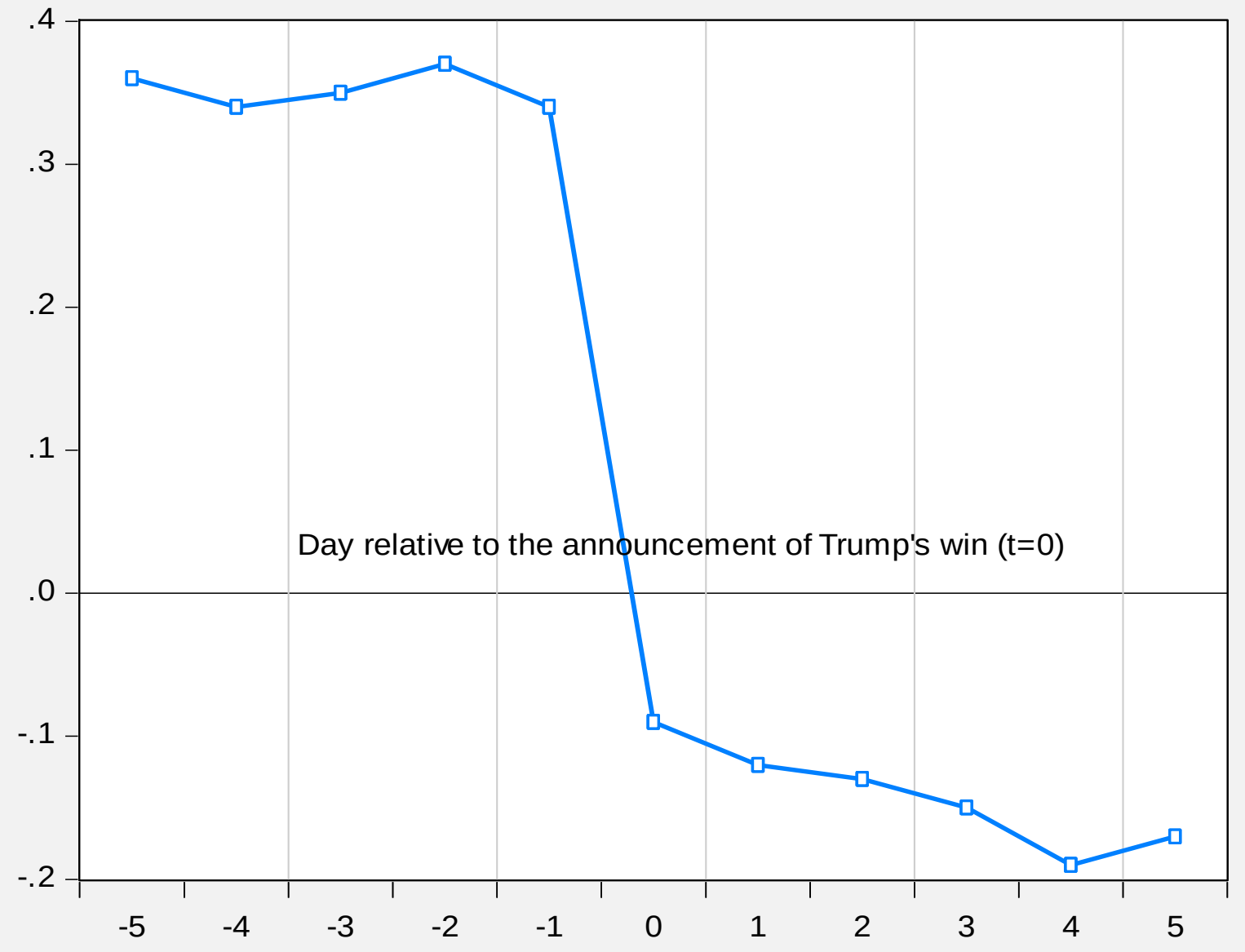

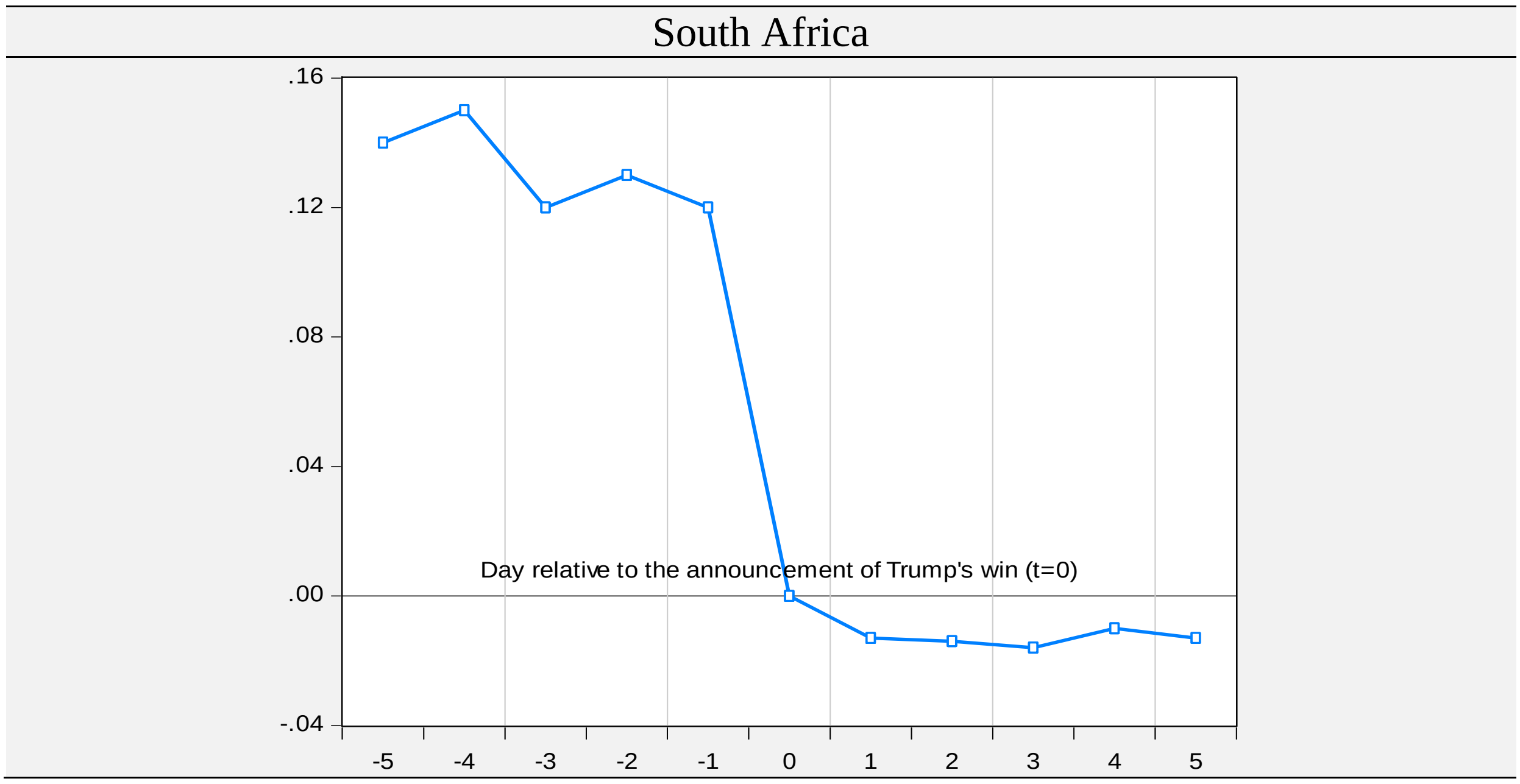

The results of the stock event study without considering potential control variables (i.e., unconditional analysis) are displayed in Table 1. We find that the announcement of Trump’s win (the event day [0; 0]) resulted in statistically significant negative CARs for all the BRICS (except Russia where we note a positive response), being somewhat stronger for China and Brazil than for India and South Africa. The BRICS-market reactions do not change in terms of sign during the [+ 1; +5] event window, but the strength of the responses appear more pronounced during the post-election period. The Russian share market, by contrast, gained markedly from this unexpected election outcome either for [0; 0] event day or over [+1; +5] event widow.

**Table 1. Trump's impacts on BRICS abnormal returns: Unconditional OLS regression results**

| | Brazil | Russia | India | China | South Africa |
|---|---|---|---|---|---|
| Event day [0 ; 0] | | | | | |
| *Constant* | 2.678432** | 3.11678** | 1.61345** | 2.13498* | 1.89742* |
| | (0.0039) | (0.0081) | (0.0072) | (0.0352) | (0.0658) |
| *Event* | **-0.09762*** ** | **0.02567*** ** | **-0.02211** ** | **-0.11435** ** | **-0.00871****** |
| | **(0.0004)** | **(0.0004)** | **(0.0014)** | **(0.0081)** | (0.0001) |
| Adjusted $R^2$ | 0.69 | 0.64 | 0.66 | 0.71 | 0.73 |
| Event window [+1; +5] | | | | | |
| *Constant* | -4.612583* | 2.96105** | 3.13492** | 1.765329 | 2.15934** |
| | (0.0355) | (0.0046) | (0.0035) | (0.1084) | (0.0023) |
| *Event* | **-0.13567*** ** | **0.099567*** | **-0.06238** ** | **-0.15673*** ** | **-0.01026*** ** |
| | **(0.0000)** | **(0.03481)** | **(0.0326)** | **(0.0002)** | **(0.0007)** |
| Adjusted $R^2$ | 0.72 | 0.77 | 0.74 | 0.70 | 0.75 |

Notes: All regressions are controlled for heteroskedasticity and the p-values are given in parentheses. *, **, *** denote statistical significance at the 10%, 5% and 1% levels, respectively.

By accounting for *WTI, Gold, Siver and Bitcoin* (Conditional analysis, Table 2), some changes with respect the strength of the Trump's victory effect (the *Event*'s coefficient become stronger by moving from the unconditional (Table 1) to the conditional analysis (Table2); this holds true over [0; 0] event day and [+1; +5] event window) were noticed. However, we usually find that the announcement of the Trump triumph in 2016 US election has varying effects across BRICS area. This event divides the BRICS equities into losers (China, Brazil, India, South Africa, in this order) and winners (Russia).The *WTI* is likely to differently affect BRICS abnormal share returns depending to whether the country is oil importer or oil exporter; while it exerted a positive effect on Russian market (exporter), its effect on the rest of BRICS (importers) stock returns seems negative. The gold and silver prices have negative influence on the abnormal cumulative returns for all the countries under study. Thus, these metals had not lost their usefulness as a safe haven to protect against deal with uncertainty over Trump's presidential win. The negative influence of Bitcoin on BRICS share returns implies that the investors in the considered countries turn to the digital currency as a refuge from weaker fiat currencies.

**Table 2. Trump's impacts on BRICS abnormal returns: Conditional OLS regression results**

| | Brazil | Russia | India | China | South Africa |
|---|---|---|---|---|---|
| **Event day [0 ; 0]** | | | | | |
| *Constant* | 1.32445*** (0.0009) | -0.026138 (0.1171) | -0.018209 (0.2281) | 0.015787 (0.1891) | 0.050083 (0.1549) |
| *Event* | **-0.133970* (0.0620)** | **0.121378** (0.0043)** | **-0.07356* (0.0339)** | **-0.193872** (0.0029)** | **-0.044113* (0.0546)** |
| *WTI* | -0.031881* (0.0202) | 0.10128*** (0.0003) | -0.01578** (0.0083) | -0.068994* (0.0304) | -0.049743** (0.0056) |
| *GOLD* | -0.023951* (0.0256) | -0.013544** (0.0073) | -0.074435* (0.0486) | -0.062891* (0.0380) | -0.04439* (0.0967) |
| *Silver* | -0.02269** (0.0035) | -0.063511* (0.0405) | -0.064773* (0.0968) | -0.074992* (0.0924) | -0.062508** (0.0043) |
| *Bitcoin* | -0.13417** (0.0015) | -0.1146* (0.0456) | -0.128721* (0.0462) | -0.19142* (0.0215) | -0.106724** (0.0095) |
| Adjusted $R^2$ | 0.89 | 0.88 | 0.93 | 0.92 | 0.91 |
| **Event window [+1; +5]** | | | | | |
| *Constant* | 1.668467* (0.0077) | 1.581424* (0.0218) | 1.26723** (0.0015) | 1.14096* (0.0456) | 1.32945 * (0.0871) |
| *Event* | **-0.169456* (0.0391)** | **0.141723** (0.0020)** | **-0.0687*** (0.0007)** | **-0.18282* (0.0367)** | **-0.069619* (0.0707)** |
| *WTI* | -0.059222** (0.0067) | 0.100776* (0.0638) | -0.08012** (0.0023) | -0.037125* (0.0282) | -0.02473** (0.0043) |
| *GOLD* | -0.059454* (0.0279) | -0.075213* (0.0955) | -0.0684*** (0.0001) | -0.110881* (0.0782) | -0.12243** (0.0079) |
| *Silver* | -0.03145* (0.0139) | -0.236306* (0.0140) | -0.064791* (0.0577) | -0.05489** (0.0096) | -0.020562* (0.0351) |
| *Bitcoin* | -0.119422 (0.3617) | -0.098422** (0.0014) | -0.143359* (0.0140) | -0.14763** (0.0064) | -0.066735* (0.0875) |
| Adjusted $R^2$ | 0.91 | 0.94 | 0.90 | 0.91 | 0.89 |

Notes: All regressions are controlled for heteroskedasticity and the p-values are given in parentheses. *, **, *** denote statistical significance at the 10%, 5% and 1% levels, respectively.

### 3.1. Regression results-based on the intention votes

Considering the intention votes through social media, search queries and public opinion polls as indicators of markets' beliefs regarding the US election (Table 3), we show that the results are still robust. In particular, *Google Trends* statistically and negatively affect Brazilian, Indian, Chinese and South African shares, while they exert a positive impact on the Russian stocks. Similar results are found when using *Twitter* hashtags and polling data (with the exception of Brazil).

**Table 3. The impacts of the intention votes on BRICS stock returns: Unconditional OLS regression results**

| | Brazil | Russia | India | China | South Africa |
|---|---|---|---|---|---|
| | | | *STR* and *Google Trends* | | |
| *Constant* | 0.763241** (0.0065) | 0.662156** (0.0059) | 0.853594* (0.0739) | 0.271307 (0.1680) | 0.00345*** (0.0007) |
| *Google Trends* | **-0.13456 (0.2451)** | **0.176446** (0.0052)** | **-0.108786* (0.0400)** | **-0.180459 (0.2558)** | **-0.01234* (0.0156)** |
| Adjusted $R^2$ | 0.83 | 0.81 | 0.82 | 0.85 | 0.83 |
| | | | *STR* and *Twitter* | | |
| *Constant* | 1.116414* (0.0425) | 1.347377* (0.0905) | 1.19710* (0.0819) | 1.565629** (0.0096) | 1.491338* (0.0315) |
| *Twitter* | **-0.168191* (0.0556)** | **0.153365** (0.0091)** | **-0.077745* (0.0806)** | **-0.14438*** (0.0001)** | **-0.085861* (0.0527)** |
| Adjusted $R^2$ | 0.89 | 0.86 | 0.85 | 0.79 | 0.85 |
| | | | *STR* and *polls* | | |
| *Constant* | 0.141563* (0.0749) | 0.175537** (0.0091) | 0.110998 (0.8754) | 0.033970 (0.1620) | 0.021178 (0.2743) |
| *polls* | **0.119329 (0.2670)** | **0.127439* (0.0425)** | **-0.07988** (0.0082)** | **-0.16188* (0.0202)** | **-0.09128*** (0.003)** |
| Adjusted $R^2$ | 0.85 | 0.82 | 0.84 | 0.88 | 0.86 |

Notes: All regressions are controlled for heteroskedasticity and the p-values are given in parentheses. *, **, *** denote statistical significance at the 10%, 5% and 1% levels, respectively.

The results of the effect of intention votes on the stock returns while considering the control variables are reported in Table 4. Whatever the public opinions proxies used (*Google Trends*, *Twitter* or *polls*), we often show that the BRICS markets are not equally vulnerable to Trump's victory. Russia appears the only winner from the US election outcome. The additional explanatory variables still exert similar effects. *WTI* impact positively the oil exporting country (Russia), while its effect on the oil importing countries seems negative which is yet highly expected. *Gold* and *Silver* affect negatively the BRICS stock returns, highlighting their viability to serve as a safe haven in this period of upheaval. Bitcoin has been shown to be negatively correlated with stock returns, pointing toward its safe haven and hedging capabilities.

Remarkably, the use of polls seems less appropriate than social media and search queries since more significant results are found for the second cases. In particular, we show that market sentiment reflected in search queries and individual text messages matters for assessing the responses of BRICS stock markets to US election event. In light of the ubiquity of social media data and the ability to deal with a large data volume, the use of this kind of data appears a quite interesting field for future studies on the effects of economic and political events. On the contrary, some polls' coefficients seem insignificant (Brazil for unconditional analysis and South Africa for conditional investigation). The polls usually report only the results while leaving out the "don't knows", and directly transform answers into opinions. Moreover, the pollsters report the beliefs of a random sample of the entire population, and thus it is not the best representative of the full public opinion. This may explain why the 2016 US election prediction were flawed. In fact, the majority of projections gave Hillary Clinton more chance of winning the US presidential election (see Appendix A). In this context, Sociologist Herbert Gans asserted that "polls are answers to questions rather than opinions". To be more effective, the pollsters should pose accurate questions, telling the politicians and the public how exactly respondents feel about such event, and if they have been politically active in behalf of these feelings.

**Table 4. The impacts of the intention votes on BRICS stock returns: Conditional OLS regression results**

| | Brazil | Russia | India | China | South Africa |
|---|---|---|---|---|---|
| *STR* and *Google Trends* | | | | | |
| *Constant* | 1.19873* (0.0200) | 1.166422* (0.0111) | 1.531872 (0.2447) | 1.133039* (0.0309) | 1.896641** (0.0030) |
| *Google Trends* | **-0.163564* (0.0621)** | **0.135711** (0.0058)** | **-0.10499* (0.0330)** | **-0.158649** (0.0025)** | **-0.098390* (0.0835)** |
| *WTI* | -0.168227* (0.0599) | 0.101875** (0.0086) | -0.050940* (0.0465) | -0.102084** (0.0014) | -0.054893* (0.0216) |
| *GOLD* | -0.092015** (0.0091) | -0.083335* (0.0116) | -0.1162*** (0.0008) | -0.142460* (0.0497) | -0.199722* (0.0343) |
| *Silver* | -0.04321*** (0.0009) | -0.03214** (0.0054) | -0.054678* (-0.0311) | -0.072341** (0.0064) | -0.034521** (0.0055) |
| *Bitcoin* | -0.10543* (-0.0674) | -0.09653** (0.0081) | -0.132452* (-0.0510) | -0.142456* (0.0431) | -0.097632* (0.0389) |
| Adjusted $R^2$ | 0.88 | 0.90 | 0.87 | 0.86 | 0.89 |
| *STR* and *Twitter* | | | | | |
| *Constant* | 1.25881** (0.0097) | 1.49428* (0.0187) | 1.53943** (0.0081) | 1.626058** (0.0017) | 1.702818* (0.0185) |
| *Twitter* | **-0.160209* (0.0616)** | **0.121423* (0.0138)** | **-0.086845* (0.0527)** | **-0.174548* (0.0019)** | **-0.09235* (0.0886)** |
| *WTI* | -0.150977* (0.0142) | 0.101423* (0.0356) | -0.027995* (0.6996) | -0.079679** (0.0011) | -0.05778*** (0.0009) |
| *GOLD* | -0.128905* (0.0474) | -0.063101** (0.0079) | -0.11304** (0.0012) | -0.101694* (0.0428) | -0.181309* (0.0556) |
| *Silver* | -0.06432** (0.0038) | -0.057234* (0.0679) | -0.0467*** (0.0000) | -0.07625** (0.0049) | -0.04693** (0.0062) |
| *Bitcoin* | -0.089972* (0.0164) | -0.069432 (0.1520) | -0.11789** (0.0013) | -0.14698*** (0.0004) | -0.08721*** (0.0009) |
| Adjusted $R^2$ | 0.87 | 0.84 | 0.88 | 0.90 | 0.92 |
| *STR* and *polls* | | | | | |
| *Constant* | 1.622108** (0.0075) | 1.60247* (0.0861) | 0.895260 (0.4508) | 1.324009 (0.2109) | -1.026138 (0.1171) |
| *polls* | **-0.162108* (0.0163)** | **0.157355** (0.0046)** | **-0.109503* (0.0286)** | **-0.183970* (0.0620)** | **0.021178 (0.2743)** |
| *WTI* | -0.050096* (0.0995) | 0.113582** (0.0029) | -0.075538* (0.0664) | -0.09188* (0.0202) | -0.071289* (0.0313) |
| *GOLD* | -0.080407* (0.0586) | -0.00919*** (0.0000) | -0.100618* (0.0603) | -0.12395* (0.0056) | -0.013544* (0.0703) |
| *Silver* | -0.034585* (0.0212) | -0.031015** 0.0018) | -0.080618* (0.0993) | -0.02266 (0.2735) | -0.063511* (0.0405) |
| *Bitcoin* | -0.134585* (0.0769) | -0.129768* (0.0187) | -0.069454* (0.0531) | -0.13417** (0.0015) | -0.1146* (0.0456) |
| Adjusted $R^2$ | 0.77 | 0.79 | 0.75 | 0.72 | 0.69 |

Notes: All regressions are controlled for heteroskedasticity and the p-values are given in parentheses. *, **, *** denote statistical significance at the 10%, 5% and 1% levels, respectively.

Using event-study methodology and the regression-based intention votes, we re-investigate the focal linkage for a restricted period that spans between 31/12/2015 and 31/12/2016. A 2SLS technique was also employed to avoid possible endogeneity bias. The results appear fairly robust to changes in time period and to the control for endogeneity problem; the same losing and winning countries were shown. To keep space, the results are available for readers upon request.

### 4.2. Interpretations

Even though the emerging markets (BRICS, particularly) haven't yet completely incorporated the economic and geopolitical implications of the Trump's agenda for the world markets, it is clearer that the BRICS stock markets are so reactive to the great uncertainty surrounding this event. The results indicate that the BRICS stock markets are not uniformly exposed to the US presidential election outcome. Trump's win divided the BRICS into highly damaged (China and Brazil), moderately threatened (India and South Africa) and benefiting (Russia) markets.

How can we explain these heterogeneous reactions of BRICS markets?

China's stock market seems the most damaged by the victory of Donald Trump. The nervousness was fueled by Trump's provocative words on the campaign trail about how China is a currency manipulator, coupled with its fierce protectionist stance much of it directed toward China. His protectionist approach could undoubtedly harm the capital and trade flows between the United States and China. It must be stressed that the United States is the largest market for Chinese exports, accounting for approximately 20 percent of the global exports. In this way, imposing a 45 percent tariff on Chinese imports into the US as Trump proclaimed in his campaign would constitute a serious risk for Chinese economy. This aggressive US trade policy could result in a substantial China's growth slow-moving coupled with a loss of manufacturing jobs. The fact that Trump's economic agenda seeks to slash China's huge trade surplus with the US would damage shares involved in Chinese exports.

Some emerging countries often indebted in greenback (including Brazil and South Africa) are heavily dependent on foreign capital. The strength of the dollar and the rise in the interest rate on the bond market are likely to prompt massive capital outflows to the United States. To this we must add that the developed countries tend to become more protectionists. The uncertainty is greater as no one knows whether the US elect-president will transform his protectionist promises into action. In any case, Trump's anti-trade rhetoric aimed at imposing a 35 and 45 percent tariffs on some products imported could be counterproductive (risk of exacerbation of currencies competition, strong appreciation of the dollar, inflation pressures, etc.). The hope for these countries is that the rise in US interest rates will be gradual. But the inflationary agenda of the new US president may force the US Central Bank to accelerate the move. Furthermore, the United States is one of South African biggest export destination and to achieve a hike in import costs will threaten South Africa's economy. However, the uncertainty arising from a Trump victory is good for the gold price, as investors turn to this yellow metal in period of upheaval. As one of the world's dominant gold producers, South Africa will benefit from the confidence in gold as a hedge or safe haven. Brazil -as commodity-dependent country- seems also poised to emerge from recession due to the surge of oil and commodity prices.

The Trump's win has also caught India's stock market off-guard. It is expected that high import tariffs would affect adversely its economy, especially with the resulted extreme volatility of its currency against dollar. But what works in India's favor is that it has relatively low external financing needs and is not largely dependent on exports, in addition to its macro-economic parameters which seem relatively stronger. This makes it insulated from the untoward shocks that may harm flows into the rest of BRICS markets.

Exceptionally, the Russian shares benefited remarkably from the announcement of Trump's victory. The positive market reaction may partly reflect hopes for sanctions against Russian companies over Crimean crisis to be eased or lifted. With Trump in the White House, Russian investors are betting that the iced US-Russia relationship may start melting due to the president-elect's affinity towards president Vladimir Putin. Besides, Trump has been keen to stimulate US commodity production, such as oil, gas and coal, so some can anticipate that the US presidential election will constrain a rise in commodity prices, benefiting the biggest energy producers like Russia.

## 5. Conclusions

Since the Trump's win in the US presidential election, analysts all over the world start asking questions on how and to what extent the uncertainty surrounding this unexpected outcome will affect the world markets, and which markets will suffer and benefit under a Trump's administration. This paper aims at offering some answers to these questions by delving into the BRICS stock markets.

Using an event-study methodology and the voter intentions in US elections-based social media, search queries and public opinion polls, we robustly find that the BRICS equities are not equally exposed to the Trump's victory. Two main groups are derived from our regression analyses; although some markets emerged losers (China, Brazil, India and South Africa, in this order), others unfolded winners (Russia).

In general, the worst-performing markets are those which (a) borrowed dollars expecting the greenback to depreciate over time; and (b) will suffer more intensely from Donald Trump's neo-mercantilist attitude and protectionist rhetoric, or more precisely its promises revolving around slapping 45 percent duty on Chinese imports into the US to make it easier for US companies to compete, stirring fears of a currency war with China and heavily punishing all companies that have sent US works overseas. Add to this the Trump's inflammatory words on the campaign trail into several issues (especially by dubbing China as a "currency manipulator") compounded investors' uncertainty across emerging markets.

The Russian market, by contrast, benefited from the unexpected US election outcome due to the Republican president's warm tone towards Putin over the campaign and the Trump's suggestions to meaningfully improve the US-Russia relationship. Regardless of the favorable reading of the US election outcome for Russian case, the country remains facing huge challenges blighting its economy such as the lack of diversification (in particular, the great dependency to volatile and speculative commodities). Under such circumstances, the ascendancy of Donald Trump to the White House as of 20 January 2017 will not be a magic bullet for the raft of Russia's serious economic problems.

Whether the US president-elect makes good on those threats, and whether his extreme rhetoric turns into actual policies, Trump's promises have varying economic and geopolitical implications. For instance, the Trump's negative stance towards China might be used politically by Chinese leaders to stoke nationalism and declare the culpability of US government rather than Chinese authorities. This is a scenario the Obama administration has been watchful to circumvent. Chinese leaders and the propaganda machines they control have yet begun using Donald Trump's to pressurize a nationalist agenda. Last but not least, Brazil, India and South Africa should also carefully assess what new geopolitical risks may emerge with the more confrontational Trump foreign policy towards countries like China or Russia, with which these countries have strong economic commitments.

## Appendix A. The US presidential election projections: A year at the polls

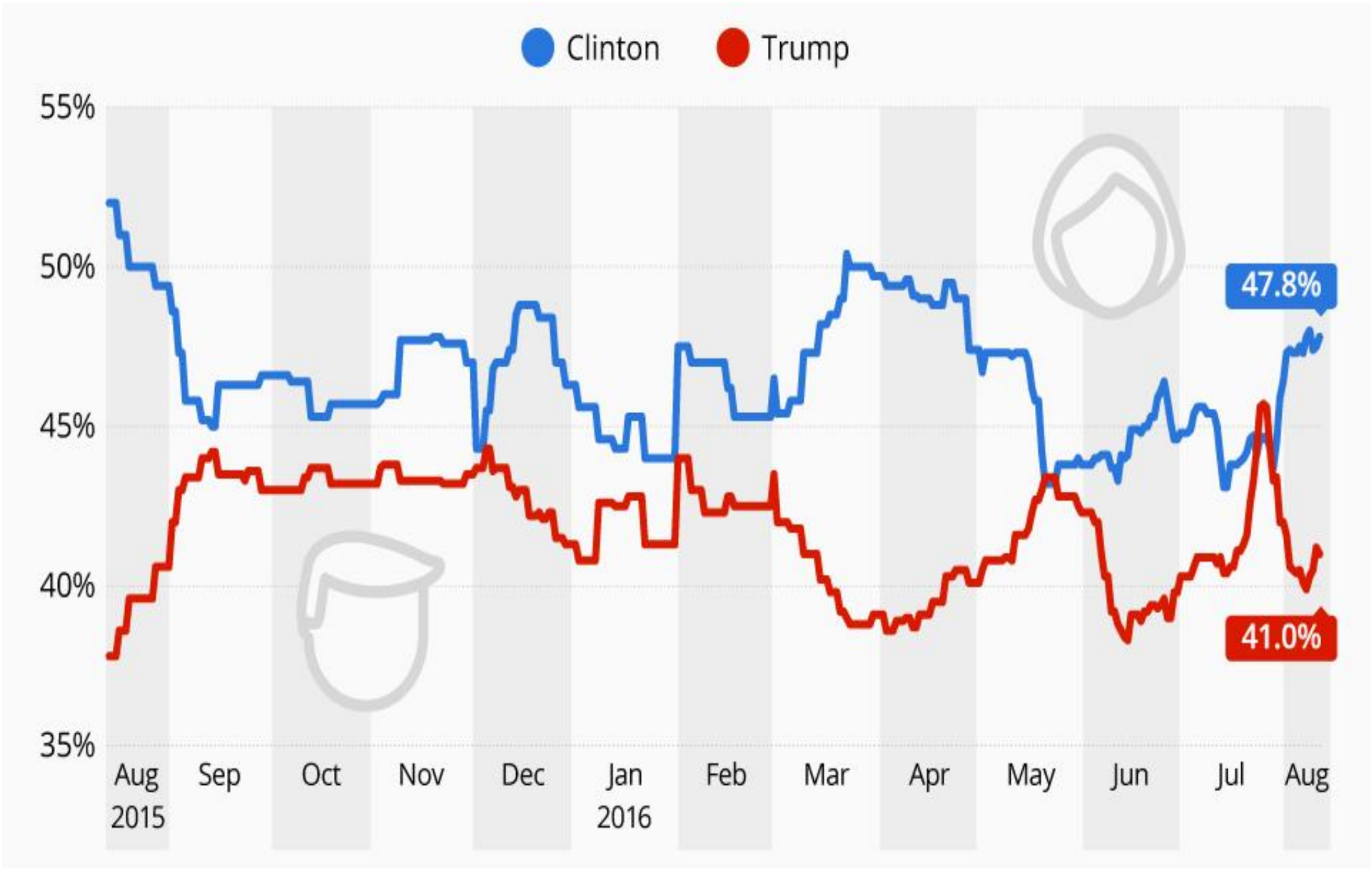

Source: Real Clear Politics.